\documentclass[final,1p,times,twocolumn]{elsarticle}

\usepackage{amssymb}

\newcommand{\Tcurie}{$T_{Curie}$}
\newcommand{\TSC}{$T_{SC}$}

\journal{Physics Research International}

\begin{document}

\begin{frontmatter}

\title{Multigap Superconductivity in the Ferromagnetic Superconductor UCoGe Revealed by Thermal Conductivity Measurements}

\author[UZH]{L.~Howald\fnref{PSI}}
\fntext[PSI]{Present address: Swiss Light Source, Paul Scherrer Institut, CH-5232 Villigen PSI, Switzerland}
\ead{ludovic.howald@psi.ch}
\author[CEA]{M.~Taupin}
\author[CEA,Toho]{D.~Aoki}

\address[UZH]{Physik-Institut der Universit\"at Z\"urich, Winterthurerstrasse 190, CH-8057 Z\"urich, Switzerland}
\address[CEA]{Institut Nanosciences et Cryog\'enie, SPSMS, CEA and University Joseph Fourier, F-38054 Grenoble, France}
\address[Toho]{IMR, Tohoku University, Oarai, Ibaraki 311-1313, Japan}

\begin{abstract}
We performed thermal conductivity measurements on a single crystal of the ferromagnetic superconductor UCoGe under magnetic field.
Two different temperature dependencies of the thermal conductivity are observed, for $\vec{H}\parallel\vec{b}$: linear at low magnetic field and quadratic for magnetic field larger than 1 Tesla. At the same field value, a plateau appears in the field dependency of the residual term of thermal conductivity. 
Such observations suggest a multigap superconductivity with a line of nodes in the superconducting gap.
\end{abstract}


\end{frontmatter}


\section{Introduction}
The orthorhombic heavy fermion system UCoGe, discovered in 2007 \cite{Huy2007}, is one of the few compounds exhibiting long range coexistence between weak itinerant ferromagnetism  (magnetic moment m$_0\cong 0.07\mu_B$ \cite{Huy2008}) and superconductivity. Such coexistence is attested by the observation of two bulk phase transitions in specific heat measurements \cite{Huy2007}. $\mu$SR and NQR measurements on different samples \cite{DeVisser2009,Ohta2010} reveal that the compound is fully ferromagnetic below the Curie temperature (\Tcurie$\cong 2.4$\,K) while about 50\,\% of the sample is superconducting below \TSC$\cong 0.5$\,K \cite{Ohta2010, Ohta2008}. 

The upper critical field is extremely anisotropic, exceeding 16 Tesla for $\vec{H}\parallel\vec{a}$ and $\vec{H}\parallel\vec{b}$ while it reaches only 0.5 Tesla for $\vec{H}\parallel\vec{c}$ \cite{Aoki2009}, the easy magnetization axis \cite{Huy2008}. The coexistence of ferromagnetism and superconductivity and the observed extremely high upper critical field suggest the realization of unconventional superconductivity with equal spin pairing (triplet) \cite{Mineev2010}. Such a superconductor is inherently two-band (one for the up and one for the down spins), however it is not known whether the two bands are superconducting or only one, in analogy to the A1 phase of $^3$He, and whether the decoupling between the bands is large enough to induce multigap superconductivity. 

Multigap superconductivity is quite common. First observed in Nb-doped SrTiO$_3$ \cite{Binnig1980}, it is found in MgB$_2$\cite{Bouquet2002}, various heavy fermions \cite{Seyfarth2005,Seyfarth2008}, cuprates \cite{Khasanov2007} and pnictides \cite{Yang2012} systems. Here we report evidences for multigap superconductivity in the ferromagnetic superconductor UCoGe.

For a two-band ferromagnetic superconductor, due to the crystal structure and the strong spin orbit coupling \cite{SamselCzekala2010}, only two types of odd parity superconducting states are possible \cite{Mineev2004}. These states differ by the position of their nodes, either lying on the northern and southern poles of the Fermi surface $k_x=k_y=0$ or on the line of equator $k_z=0$. The symmetry of the superconducting gap has a small effect on the shape of the upper critical field, which was used by Hardy and Huxley \cite{Hardy2005} to suggest a superconducting state with a line of nodes in the parent system URhGe. However, in contradiction to the theoretical prediction this line was proposed to occur at $k_x=0$. Thermal conductivity is a strong probe of the gap symmetry of a superconductor. Indeed, the density of non-superconducting quasiparticles, which are the main heat carrier channel at low temperatures, is strongly influenced by the presence and type of gap nodes. 

\section{Method and Raw Data}
\begin{figure}[hbt]
\begin{center}
\includegraphics[width=0.48\textwidth]{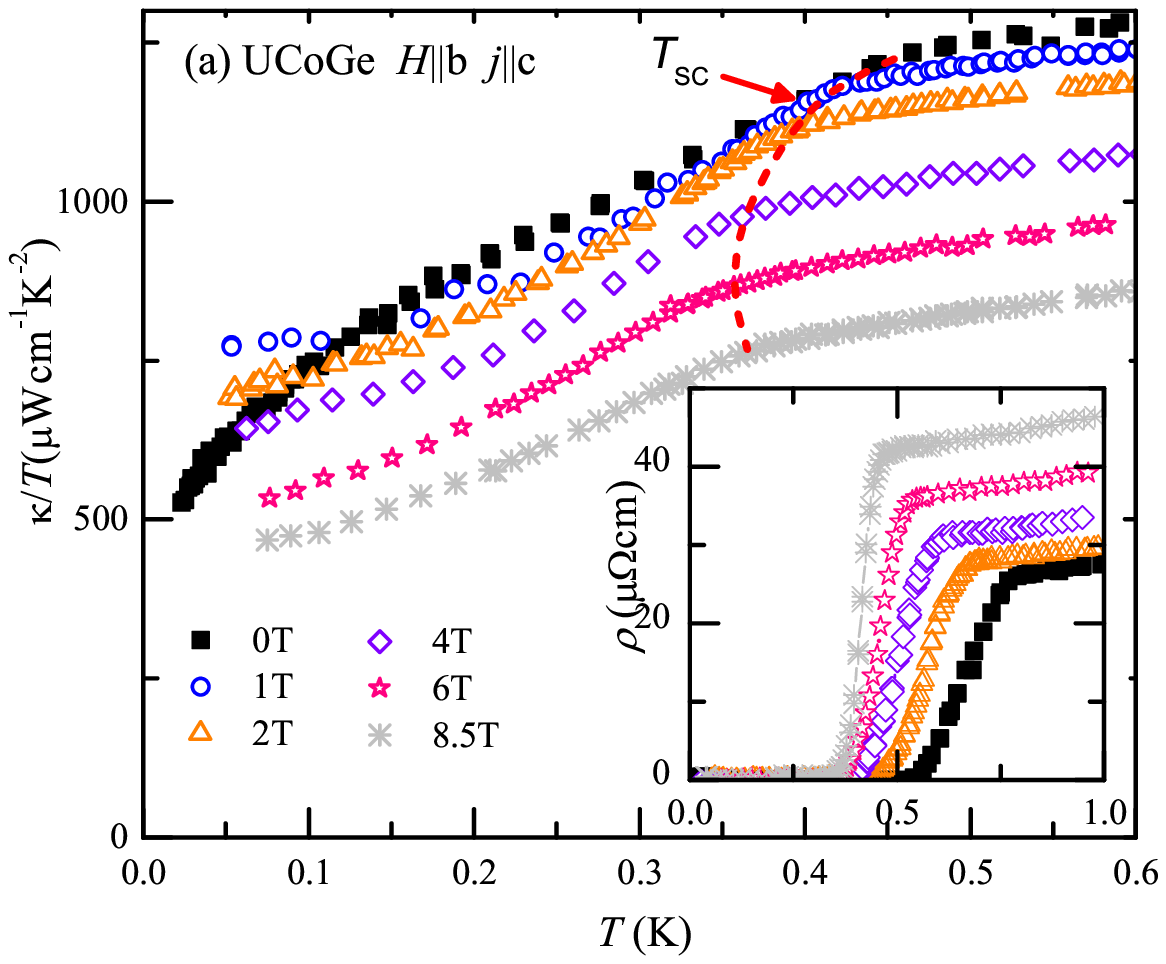}% Here is how to import EPS art
\includegraphics[width=0.48\textwidth]{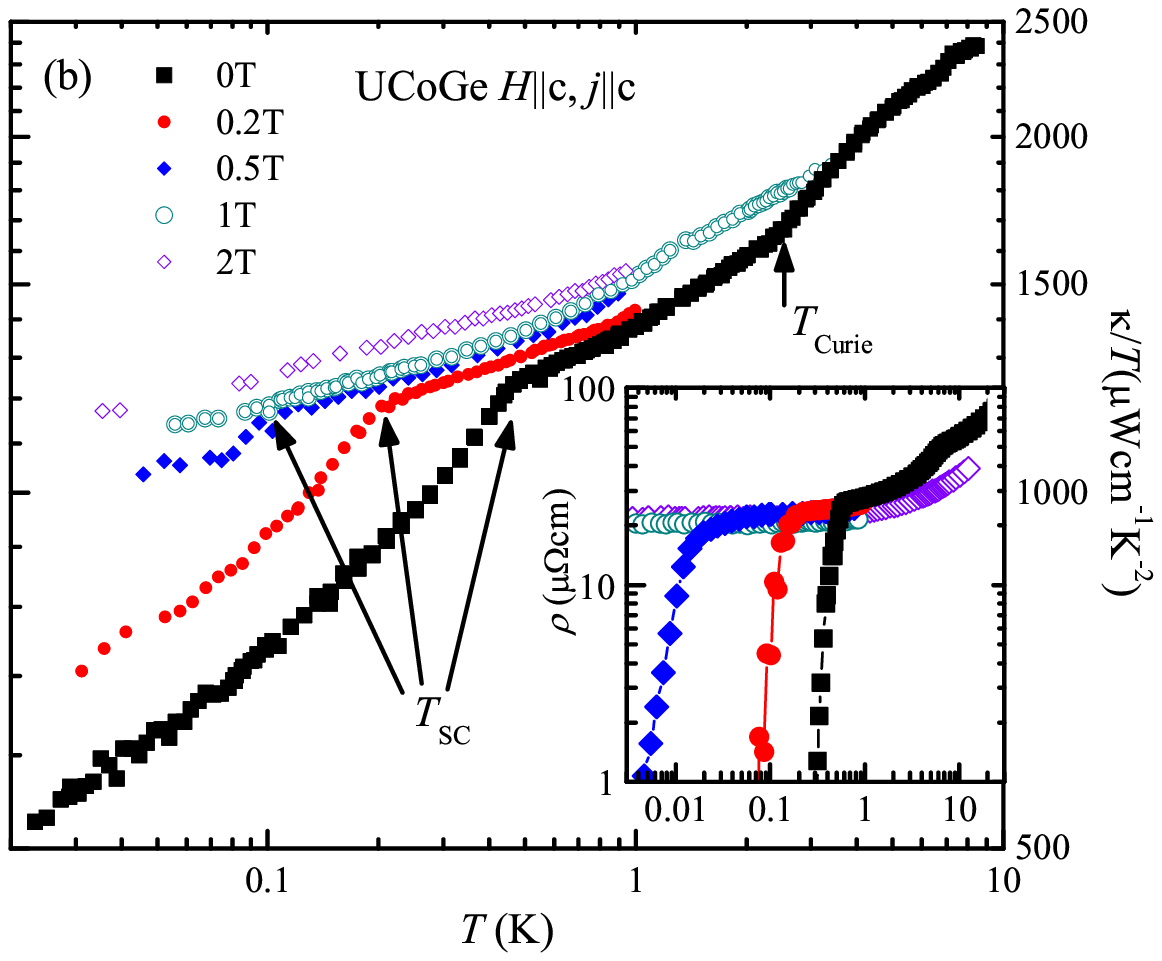}% Here is how to import EPS art
\caption{\label{fig:Kappa_rho_Hb} (color online) Raw data for thermal conductivity divided by temperature and electrical resistivity in inset, with  (a) $\vec{H}\parallel\vec{b}$-axis and (b) $\vec{H}\parallel\vec{c}$-axis. Heat current is applied along the $\vec{c}$ crystallographic axis.}
\end{center}
\end{figure}

We measured thermal conductivity ($\kappa$) with a two thermometers one heater setup in the temperature range 30\,mK -- 10\,K and in magnetic field up to 8.5 Tesla. Resistive carbon thermometers were used, held by thin Kevlar strings and measured through superconducting NbTi wires to insure good external thermal insulation. Thermometers and sample were connected through a gold wire spot welded on the sample side, achieving a contact resistance of $\sim15$\,${\mu}{\Omega}$. Four probes electrical resistivity ($\rho$) was measured simultaneously using the same gold wires for voltage measurement, allowing a direct verification of the setup using the Wiedemann--Franz law ($\kappa \rho / T \rightarrow L_0$ for $T\rightarrow 0$ with $L_0=2.44\cdot 10^{-8}$ W$\Omega$K$^{-2}$).  This setup was mounted on a piezo rotator which allows fine tuning of the relative angle between magnetic field and sample crystallographic axis. This is required due to the strong angular dependence of magnetic properties \cite{Aoki2009}. The single crystal of pure UCoGe composition was grown using the Czochralski method in a tetra--arc furnace, characterized by Laue diffraction and specific heat. The sample was cut in a bar shape with the widest direction of $l\simeq 2$\,mm along the $\vec{c}$ crystallographic axes. 
With a residual resistivity ratio of $RRR\cong 16$, the sample offers a good compromise between crystallographic quality (single grain crystal) and low mean free path. The latter property is required in order to differentiate between electronic and magnetic thermal conductivity contributions as described below.

Thermal conductivity divided by temperature ($\kappa/T$) and resistivity ($\rho$) for two field orientations, $\vec{H}\parallel\vec{b}$-axis and $\vec{H}\parallel\vec{c}$-axis, are presented in figures \ref{fig:Kappa_rho_Hb} (a) and (b) respectively. The ferromagnetic transition (\Tcurie ) disappears with a magnetic field applied along the $\vec{c}$ crystallographic axis in agreement with such an orientation for the magnetic moments, due to the absence of symmetry breaking at the transition.
The superconducting transition (\TSC ) is clearly observed by a kink in the two sets of curves below $\sim 0.5$\,K. 
The observation of both superconducting and ferromagnetic transitions in thermal conductivity curves indicates that both phases are bulk. The enhancement of \TSC{} under magnetic field ($\vec{H}\parallel\vec{b}$) previously observed in resistivity measurements \cite{Aoki2009} is confirmed as a bulk property (red dashed line in figure~\ref{fig:Kappa_rho_Hb}(a)) which indicates a high precision in the sample alignment. Indeed, such enhancement was reported to occur only when the magnetic field is applied within 1$^\circ$ of the $\vec{b}$ crystallographic axis \cite{Aoki2011b}. It is particularly interesting to note that the bulk re-entrance of superconductivity already occurs at $\sim$5\,Tesla, while the resistive one is only observed around 10\,Tesla \cite{Aoki2009}. The method used to extract \TSC{} is described in \cite{HowaldThese}.
The residual value in the superconducting state is quite large ($\sim 50$\% of the normal state value).
For a fully open superconducting gap (s-wave) the ratio $\kappa/T$ vanishes with temperature ($\lim_{T\rightarrow 0}\kappa/T=0$), as observed in Nb\cite{Lowell1970}, NbSe\cite{Boaknin2003} and MgB$_2$\cite{Sologubenko2002} for example. For systems in which the superconducting gap vanishes at points or lines of nodes, a residual value is expected due to impurity scattering. Such residual value was observed in the d-wave high-\TSC{} superconductor Tl$_2$Ba$_2$CuO$_{6+\delta}$ \cite{Proust2002} and in the possibly p-wave superconductor Sr$_2$RuO$_4$ \cite{Suderow1998,Suzuki2002}. For systems in which a superconducting gap with lines of nodes is present, the limit can even be universal, with $\left(\kappa(T)T_{SC}/(\kappa(T_{SC})T)\right)_{T\rightarrow 0}=C$ independent of the amount of impurities \cite{Graf1996}.

In UCoGe, the residual value of thermal conductivity is not universal. A recently probed sample with RRR$\cong100$ has a lower residual term of $\sim30$\% of the extrapolated non superconducting value \cite{Taupin2014}. The residual term can be due to the presence of nodes in the superconducting gap, a special superconducting phase (like superfluid phase A1 of $^3$He), a band of gapless superconductivity (due to impurities) or an inhomogeneous sample (partly non superconducting). This last option is supported by NQR measurements, reporting a mixture of superconducting and non-superconducting regions \cite{Ohta2010}, and by a specific heat measurement reporting a residual term of about 50\% of the normal state value \cite{Aoki2012}. Both experiments were performed on samples of similar quality to the one used in this study. Note that the sharpness of both the ferromagnetic and the superconducting transitions exclude the possibility of a distribution of transitions in the bulk phases, which would give rise to broad or no features in thermal conductivity. The higher value for \TSC{} observed in the resistivity measurement must originate from part of the sample with non bulk superconductivity. This could be due to filamentary superconductivity or superconductivity occurring first in ferromagnetic domain wall, where magnetization is suppressed which is certainly favorable for superconductivity.

\section{Results and Discussion}
\begin{figure}[htb]
\begin{center}
\includegraphics[width=0.48\textwidth]{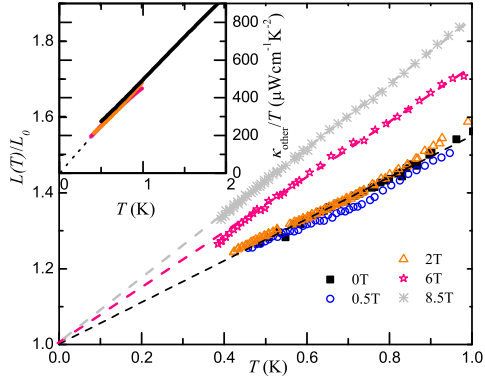}% Here is how to import EPS art
\caption{\label{fig:LL0} (color online) Lorenz ratio $\kappa\rho/(L_0 T)$ obtained from the thermal conductivity and resistivity data presented in figure \ref{fig:Kappa_rho_Hb}a. All the curves extrapolate to 1 at $T=0$\,K, reflecting the good quality of the measurement. In inset the residual thermal conductivity contribution ($\kappa_{other}/T$) is shown (see text).}
\end{center}
\end{figure}

\begin{figure*}[htb!]
\begin{center}
\includegraphics[width=0.95\textwidth]{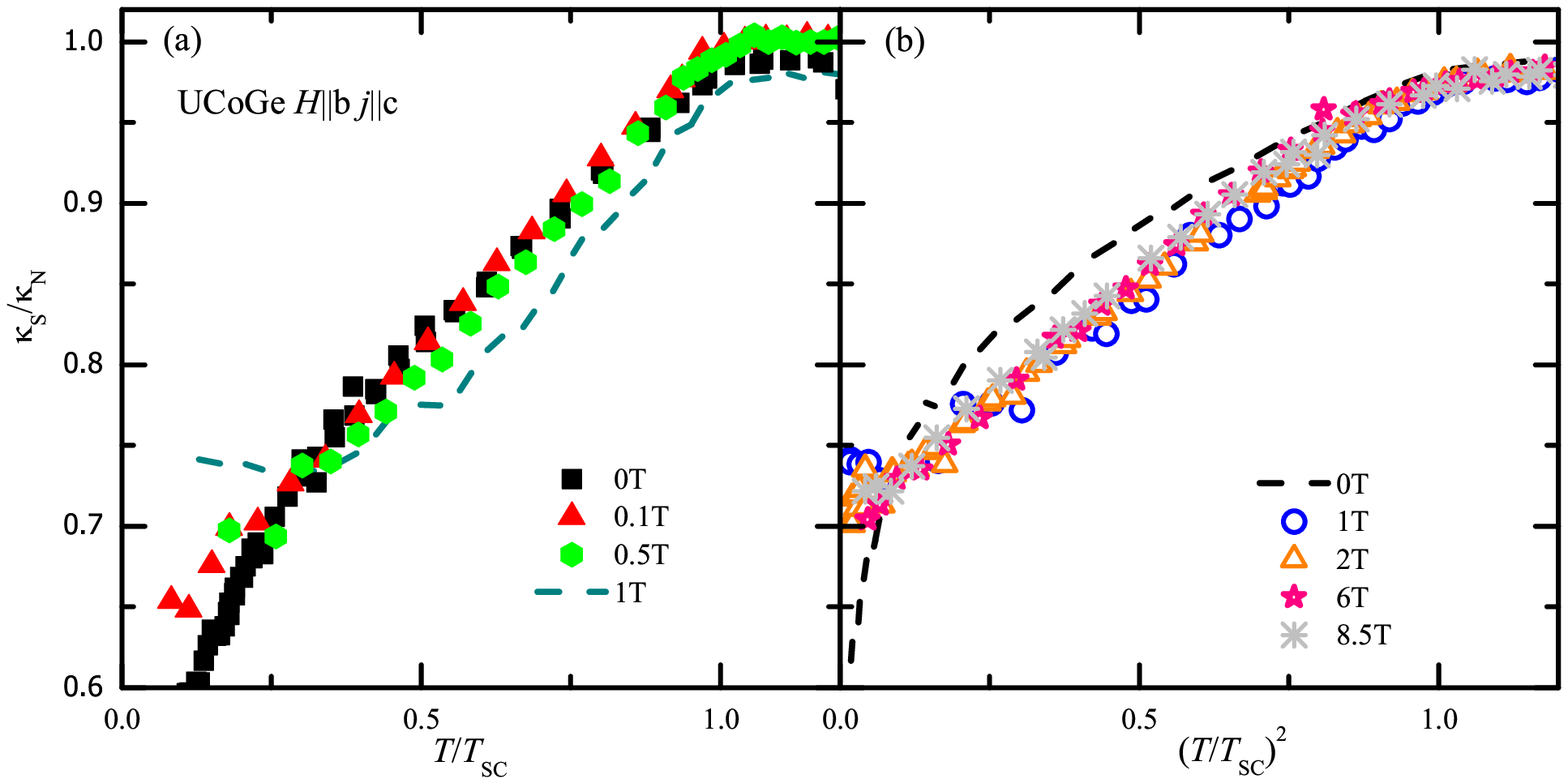}\\ % Here is how to import EPS art
\includegraphics[width=0.48 \textwidth]{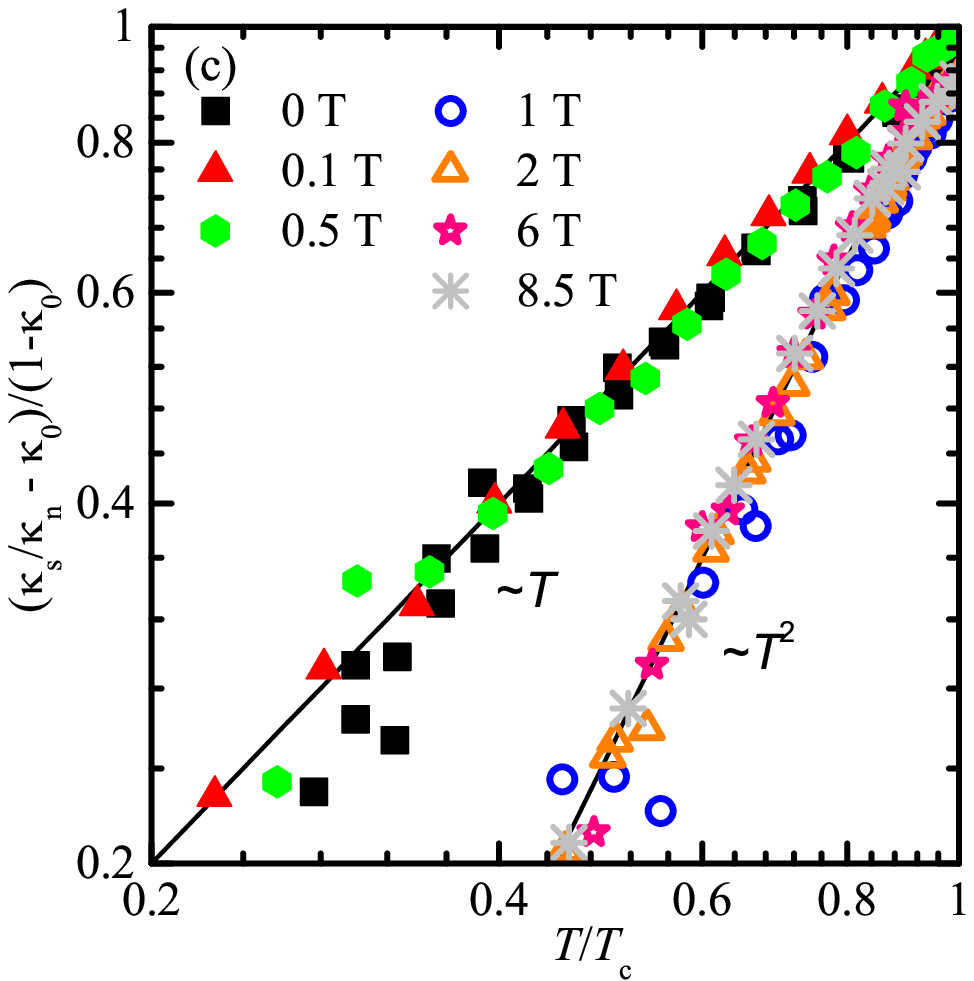}
\caption{\label{fig:Ks_Kn_T} (color online) Ratio of the superconducting to normal state value of the electronic contribution to thermal conductivity. (a) This ratio is linear in temperature at low fields $H<1$\,T, quadratic in temperature for higher fields (b). (c) The two temperature dependencies are best viewed in a double logarithmic plot after subtraction of the residual term $\kappa_0$. $\kappa_0=\lim_{T\rightarrow 0}\kappa_S/\kappa_N(T)$ was linearly extrapolated from panels (a) \& (b) in the temperature ranges $0.25<T/T_c<0.97$ and $0.08<(T/T_c)^2<0.97$ respectively.}
\end{center}
\end{figure*}

In figure~\ref{fig:LL0} the Lorenz ratio ($L/L_0=\kappa\rho/(T L_0)$) calculated for different values of the magnetic field applied along the $\vec{b}$ crystallographic direction is plotted.
For $T\rightarrow 0$ the Wiedemann-Franz law $L=L_0$ is obeyed within 5\% which reflect the validity of our measurements. The equality indicates that at low temperatures, electrical and thermal currents are transported by the same carriers: the electrons. At finite temperatures two effects can produce a deviation from the unity of $L/L_0$. Other thermal carrier channels such as phonon or magnon excitations will enhance this ratio. In contrary the ratio will be lower than one if a strong electron--electron inelastic scattering, reducing the efficiency of electronic thermal conductivity, dominates \cite{Ziman1979}. In a conventional metal, magnons are absent and as few phonons are present at low temperatures, the ratio is reduced below 1 for the lowest temperatures and exceeds 1 in the higer temperatures regime due to the phonons contribution. The depth of the minimum in $L/L_0$ depends on the mean free path of the quasiparticles. 

As the sample investigated has a relatively low $RRR$ value the mean free path of the quasiparticles is short and only a weak deviation from 1 is expected for $L/L_0$ at low temperatures.
In figure \ref{fig:LL0} the increase of $L/L_0$ above 1 from the lowest temperatures indicates the presence of another type of heat carrier than the electrons. As UCoGe is ferromagnetic, magnons or uniaxial fluctuations are good candidates \cite{Ohta2008,HowaldThese,Prokes2010}. An analysis of the anisotropy of this contribution with heat current direction re-enforce this hypothesis \cite{Taupin2014}. We can obtain the approximate value of this additional thermal conductivity contribution ($\kappa_{other}$) by assuming the Lorenz ratio is 1 for the electronic contribution: $L=(\kappa_{el}+\kappa_{other})\rho/T$ with $L_0=\kappa_{el}\rho/T$ $\rightarrow$ $\kappa_{other}/T=(L-L_0)/\rho$. We found that the magnetic contribution is independent of a magnetic field applied along the $\vec{b}$ crystallographic direction (inset figure \ref{fig:LL0}) while it is strongly reduced by a magnetic field applied in the $\vec{c}$ crystallographic direction, as expected for longitudinal spin fluctuations \cite{HowaldThese} and in agreement with a previous study using a different technique to extract the magnetic contribution \cite{Taupin2014}. 

In order to further analyze the temperature dependence of thermal conductivity, we subtract the additional contribution obtained previously ($\kappa_S=\kappa-\kappa_{other}$) and calculated the thermal conductivity one would observe if the compound was not superconducting, hereafter called normal contribution to thermal conductivity ($\kappa_{N}$). 
This is done by extrapolating the normal state resistivity to $T\rightarrow 0$ assuming a Fermi liquid dependence ($\rho(T)=\rho_0+AT^2$).  Then we calculated the electronic contribution to thermal conductivity using the Wiedemann--Franz law:  $\kappa_{N}/T=L_0/\rho$. Note that this is only possible due to the moderate $RRR$ value of the sample, when practically no deviation from the Wiedemann--Franz law are expected. Indeed, for a more general study \cite{Taupin2014}, a phenomenological model had to be introduced for the electronic thermal conductivity which prevent any further discussion of the temperature dependence.

The ratio of superconducting to ``extrapolated normal state'' thermal conductivity in the superconducting state ($\kappa_S/\kappa_N$) is reported in figure \ref{fig:Ks_Kn_T} (a) and (b) for different magnetic fields applied along the $\vec{b}$ crystallographic direction. Such ratio is related to the fraction of superfluid quasiparticles. We can clearly distinguish two different temperatures dependencies: linear for $\mu_0 H < 1$\,T and quadratic otherwise. Panel (c) of figure~\ref{fig:Ks_Kn_T} emphases the linear and quadratic temperature dependencies with a double logarithmic plot. %The small deviations from linearity for $\mu_0H = 0.2 0.5$ and 1 T result from the correlations between the residual term and the exponent and suggest a small crossover region. 
The temperature dependence of thermal conductivity is related to the type of nodes (points, lines) and their opening angle \cite{Matsuda2006}. The different temperature dependencies indicate two different gap structures depending on the field range.  There is no report of a phase transition between two different superconducting phases, neither with temperature nor upon applying a magnetic field, as required in order to modify the symmetry of the superconducting gap and our measurements support a crossover. Therefore, we infer a multigap superconducting state and not multiple superconducting states. The low field temperature dependence would then result from the addition of the thermal conductivity of the two bands, while at high fields only the band with the larger superconducting gap would be superconducting.  The high magnetic field temperature dependence, $\kappa_S/\kappa_N \sim T^2$, suggests the presence of a line of nodes in the gap of the superconducting band, as expected in analogy to URhGe \cite{Hardy2005}. Note that the observation of two different field ranges is independent of the temperature dependence of the subtracted $\kappa_{other}$ contribution. A similar evolution of the temperature dependence of thermal conductivity was observed in well--known two-band superconductors such as MgB$_2$ \cite{Sologubenko2002}, CeCoIn$_5$ \cite{Seyfarth2008} and NbSe$_2$ \cite{Boaknin2003} although with different power laws. The small deviation from linearity of $\kappa_S/\kappa_N$ at low temperatures ($T/T_{SC}<0.25$) for $\mu_0H=0$\,T is understood in the multigap scenario as corresponding to the characteristic energy of the smaller gap (figures \ref{fig:Kappa_rho_Hb} (a) and \ref{fig:Ks_Kn_T}).

\begin{figure}[htb!]
\begin{center}
\includegraphics[width=0.48\textwidth]{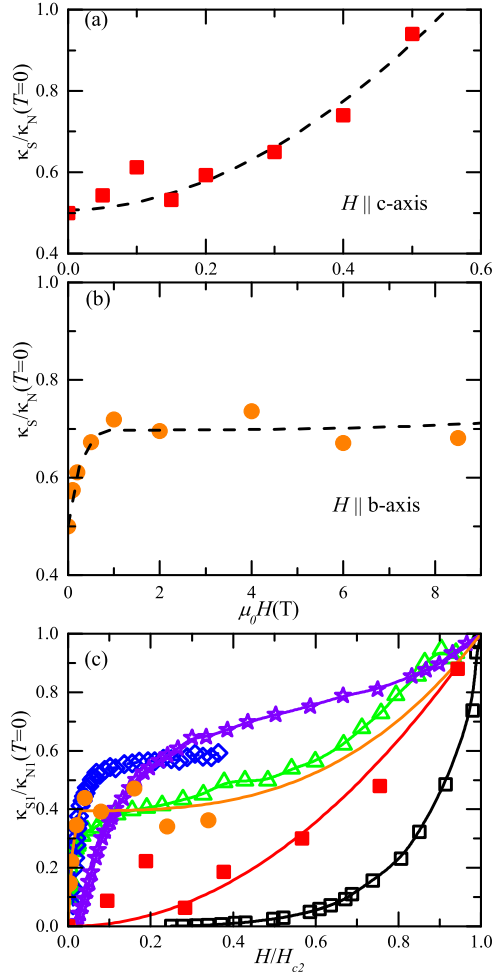}% Here is how to import EPS art
\caption{\label{fig:Ks_Kn_H} (color online) Field dependence of the ratio $\kappa_S(H)/\kappa_N(H)$: (a) for $\vec{H}\parallel\vec{c}$ (red squares), (b) for $\vec{H}\parallel\vec{b}$ (orange circles). (c) Comparison between $\kappa_{S1}(H)/\kappa_{N1}(H)$ (see text) of UCoGe and the single gap superconductor Nb (black squares \cite{Lowell1970}) and the two-band superconductors PrOs$_4$Sb$_{12}$ (green triangles \cite{Seyfarth2005}), MgB$_2$ $\vec{H}\parallel(a,b)$ (blue diamonds \cite{Sologubenko2002}) and MgB$_2$ $\vec{H}\parallel\vec{c}$ (violet stars \cite{Sologubenko2002}). $H_{c2}\parallel\vec{b}$ of UCoGe is taken as 25 Tesla in (c).}
\end{center}
\end{figure}

The idea of a multigap superconductivity is re-enforced by the appearance of a plateau above $\mu_0H>1$\,T in the field evolution of the residual term of thermal conductivity, for $\vec{H}\parallel\vec{b}$ (figure \ref{fig:Ks_Kn_H} (b)). The experimental resolution might not allow to observe such feature for $\vec{H}\parallel\vec{c}$ (figure \ref{fig:Ks_Kn_H} (a)). The decoupling between the two gaps might also be weaker in this configuration.
If we assume that the residual term for $\mu_0H=0$\,T is due to an inhomogeneous part of the sample never superconducting, we can extend the analysis by comparing UCoGe 
to well established two gaps superconductors, as PrOs$_4$Sb$_{12}$ and MgB$_2$ (Fig. \ref{fig:Ks_Kn_H} (c)). Here we have assumed two parallel contributions to thermal conductivity $\kappa_i=\kappa_{i1}+\kappa_2$, with $(i=S,N)$ and $\kappa_2$ for the never superconducting contribution. $\kappa_2$ is assumed to be field independent. 
Independently of this assumption, the three systems: UCoGe, PrOs$_4$Sb$_{12}$ and MgB$_2$ presented in figure \ref{fig:Ks_Kn_H} are characterized by two energy scales corresponding to the values of the two respective gaps. The field dependence of $\kappa_{S}/\kappa_{N}$ for a single gap superconductor is drastically different to the one found in UCoGe as demonstrated with the case of Nb. 

A ferromagnetic system has inherently two bands, for majority and minority electron spins. It is therefore tempting to map the two superconducting gaps to the two ferromagnetic bands.
The different strength of superconductivity could then be explained by the proximity to the Lifshitz phase transition, previously reported \cite{Malone2012}. 
The strong increase of density of state would enhance the superconducting coupling. It is however not clear whether such bands could be decoupled enough electronically to induced multigap superconductivity. A classical scenario of the two gaps occurring on different Fermi pockets is another possibility.

\section{Conclusions}
In conclusion, with the study of thermal conductivity ($\kappa(T,H)$) in the ferromagnetic heavy fermion system UCoGe, we confirm the re-enforcement of superconductivity under magnetic field and establish the bulk character of this effect. In addition, we observed two distinct energy scales depending on the value of the magnetic field applied along the $\vec{b}$ crystallographic axis. At high magnetic fields the temperature dependence of thermal conductivity is compatible with the presence of a line of nodes in the superconducting gap as reported in the parent system URhGe. Both the field dependence of the residual term and the different temperature dependencies of $\kappa(T,H)$ suggest the realization of multigap superconductivity. 
Further experiments on samples with different $RRR$ values and heat currents directions as well as theoretical modeling of thermal conductivity in multigap systems are required in order to obtain the exact nature of the superconducting state in the ferromagnetic superconductor UCoGe.

\section*{Acknowledgments}
We gratefully acknowledged experimental support from J.P.~Brison and sample preparation from V.~Taufour. This work was supported by the Commissariat \`a l'\'energie atomique of Grenoble and by the Universit\'e de Grenoble.

\end{document}